\documentclass[twocolumn, journal]{IEEEtran}

\IEEEoverridecommandlockouts

\usepackage{bm}
\usepackage{color}
\usepackage{amsthm}
\usepackage{amsmath}
\usepackage{amssymb}
\usepackage{amsfonts}
\usepackage{graphicx}
\usepackage{algorithm}
\usepackage{algorithmic}
\usepackage{ulem}
\usepackage{lineno}
\usepackage{lettrine}
\usepackage{subfigure}
\usepackage{epstopdf}
\usepackage{stfloats}
\usepackage{multirow}
\usepackage{booktabs}
\usepackage{array}
\usepackage{setspace}
\usepackage{makecell}
\usepackage{textcomp}
\usepackage{xcolor}
\usepackage{caption}

\newcommand{\be}{\begin{equation}}
\newcommand{\ee}{\end{equation}}
\newcommand{\bea}{\begin{eqnarray}}
\newcommand{\eea}{\end{eqnarray}}
\newcommand{\ba}{\begin{array}}
\newcommand{\ea}{\end{array}}
\newcommand{\nid}{\noindent}

\title{Partially Distributed Beamforming Design for RIS-Aided Cell-Free Networks
\thanks{P. Ni, M. Li, and R. Liu are with the School of Information and Communication Engineering, Dalian University of Technology, Dalian 116024, China (e-mail: pfni@mail.dlut.edu.cn; mli@dlut.edu.cn; liurang@mail.dlut.edu.cn).}
\thanks{Q. Liu is with the School of Computer Science and Technology, Dalian University of Technology, Dalian 116024, China (e-mail: qianliu@dlut.edu.cn).}
}

\author{Pengfei Ni,
        Ming Li,~\IEEEmembership{Senior Member,~IEEE,}
        Rang Liu,~\IEEEmembership{Graduate Student Member,~IEEE,}\\
        and Qian Liu,~\IEEEmembership{Member,~IEEE}
        \vspace{-0.6 cm}}

\begin{document}

\maketitle
\pagestyle{empty}
\thispagestyle{empty}
\begin{abstract}
Cell-free networks are regarded as a promising technology to meet higher rate requirements for beyond fifth-generation (5G) communications.
Most works on cell-free networks focus on either fully centralized beamforming to maximally enhance system performance, or fully distributed beamforming to avoid extensive channel state information (CSI) exchange among access points (APs).
In order to achieve both network capacity improvement and CSI exchange reduction, we propose a partially distributed beamforming design algorithm for reconfigurable intelligent surface (RIS)-aided cell-free networks.
We aim at maximizing the weighted sum-rate of all users by designing active and passive beamforming subject to transmit power constraints of APs and unit-modulus constraints of RIS elements.
The weighted sum-rate maximization problem is first transformed into an equivalent weighted sum-mean-square-error (sum-MSE) minimization problem, and then alternating optimization (AO) approach is adopted to iteratively design active and passive beamformer.
Specifically, active beamforming vectors are obtained by local APs and passive beamforming vector is optimized by central processing unit (CPU).
Numerical results not only illustrate the proposed partially distributed algorithm achieves the remarkable performance improvement compared with conventional local beamforming methods, but also further show the considerable potential of deploying RIS in cell-free networks.
\end{abstract}


\section{Introduction}
Cell-free networks have been regarded as a promising technology for beyond fifth-generation (5G) communications owing to its benefits of providing higher data rates, more uniform coverage, and better ability to manage interference \cite{E. Bjornson2017}.
In a cell-free network, plenty of distributed access points (APs) that are connected to a central processing unit (CPU) simultaneously serve all users via time-division duplex (TDD) mode \cite{H. Q. Ngo2017}, \cite{O. T. Demir2021}.
To take full advantages of cell-free networks, beamforming techniques are very imperative.
Some fully distributed beamforming approaches have been advocated since they can avoid extensive channel state information (CSI) exchange \cite{Y. Zhang2019}.
However, multi-user interference cannot be efficiently eliminated without the collaboration among APs.
Recently, a comprehensive analysis illustrated that the higher level of cooperation can provide significant performance improvement \cite{E. Bjornson2020}.
Unfortunately, with the increasing number of APs and users, it is impractical to accomplish fully centralized beamforming design for CPU to collect all instantaneous CSI from APs.
Hence, partially distributed or cooperative distributed beamforming design was proposed to make better trade-off between network capacity and computational complexity \cite{I. Atzeni2021}.

On the other hand, reconfigurable intelligent surface (RIS)-aided communications have attracted significant attention in recent years \cite{Q. Wu2019}-\cite{C. Pan2021}.
In \cite{Z. Zhang2021}, a centralized joint precoding framework was proposed to improve network capacity for wideband RIS-aided cell-free networks.
In \cite{S. Huang2021}, the authors developed a fully decentralized cooperative beamforming design framework based on alternating direction method of multipliers (ADMM).
Although less backhaul signaling is needed for local variables updating in each iteration, the convergence of ADMM algorithm requires many iterations so that frequent signaling exchange among APs is required and it is inefficient.
In addition, optimizing a large-dimensional passive beamforming in local APs not only increases the deployment cost of processor per AP, but also leads to more latency time for distributed beamforming schemes.

In order to achieve network capacity improvement, we introduce RIS into cell-free networks and focus on maximizing the weighted sum-rate by designing active beamformer and passive beamformer subject to transmit power constraints at APs and unit-modulus constraints of RIS elements.
The weighted sum-rate maximization problem is first transformed into an equivalent weighted sum-mean-square error (sum-MSE) minimization problem and then alternating optimization (AO) approach is adopted to iteratively optimize active beamformer and passive beamformer.
To avoid CSI exchange between APs and take full advantage of the centralized processing ability of CPU, we propose a partially distributed beamforming algorithm, in which the CPU takes responsibility for large-dimensional passive beamforming optimization and each AP only needs to locally optimize its small-dimensional active beamformer.
Then, passive beamforming vector is optimized by CPU with strong processing capability.
Finally, we provide numerical simulation results to illustrate the effectiveness of proposed partially distributed beamforming design algorithm as well as demonstrate the significant potential of deploying RIS in cell-free networks.

\section{System Model and Problem Formulation}
\begin{figure}[t]
\centering
  \includegraphics[width = 2.6 in]{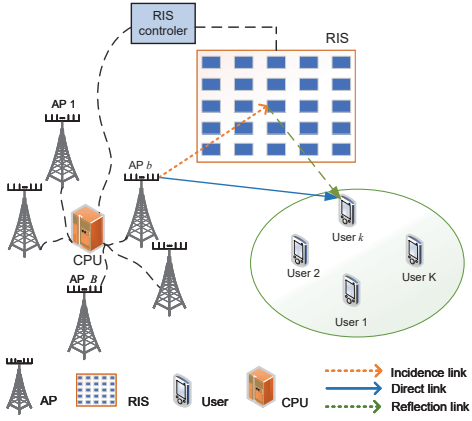}
  \caption{RIS-aided cell-free networks.}
  \label{fig:system model}
  \vspace{-0.2 cm}
\end{figure}

We consider a downlink RIS-aided cell-free network, wherein a set of multi-antenna APs $\mathcal{B} \triangleq \{1, \ldots, B\}$  simultaneously serves a set of single-antenna users $\mathcal{K} \triangleq \{1, \ldots, K\}$ with the aid of one RIS, which is controlled by APs or CPU via wired or wireless links, as shown in Fig. \ref{fig:system model}.
Each AP is equipped with $N_{\mathrm{t}}$ antennas and RIS consists of $M$ reflecting elements.
The APs are connected to a CPU through the fronthaul links.
The transmission from APs to users are operated by TDD mode, including two phases for each coherence interval: (\textit{i}) Uplink training; (\textit{ii}) Downlink data transmission.

\subsection{Uplink Training and Channel Model}
In the uplink training phase, all $K$ users simultaneously send pilot sequences to APs and each AP estimates CSIs via the received signal.
To be specific, let $\mathbf{h}_{b,k} \in \mathbb{C}^{N_{\mathrm{t}}}$, $\mathbf{G}_{b} \in \mathbb{C}^{M\times N_{\mathrm{t}}}$, and $\mathbf{v}_{k} \in \mathbb{C}^{M}$ denote the downlink channel from AP-$b$ to user-$k$, from AP-$b$ to RIS, and from RIS to the $k$-th user, respectively.
After the uplink training phase, AP-$b$ obtains the channels $\mathbf{h}_{b,k}$, $\mathbf{G}_{b}$, and $\mathbf{v}_{k},\forall k$. Note that the channels from RIS to users (i.e., $\mathbf{v}_{k},\forall k$) are both known for CPU and APs.

With the reflected channel via RIS, the equivalent baseband channel between AP-$b$ and user-$k$ is modeled as
\begin{equation}
\begin{aligned}\label{eq:equivalent channel}
\tilde{\mathbf{h}}_{b,k}^H &= \mathbf{h}_{b,k}^H  + \mathbf{v}_{k}^H \bm{\Phi} \mathbf{G}_{b},=  \mathbf{h}_{b,k}^H  + \bm{\theta}^H \mathrm{diag}(\mathbf{v}_{k}^H) \mathbf{G}_{b},
\end{aligned}
\end{equation}
where $\bm{\Phi} \triangleq \mathrm{diag}(\phi_1, \ldots, \phi_m) \in \mathbb{C}^{M \times M}$ denotes the phase-shift matrix of RIS.
By introducing an auxiliary vector $\bm{\theta}$ and then based on the relationship, i.e., $\theta_m = (\phi_m)^*, \forall m$, the reformed phase-shift vector can be expressed as $\bm{\theta}^H  \triangleq [\theta_1,\ldots,\theta_m] \in \mathbb{C}^{1\times M}$, where $|\theta_m| = 1, \forall m$.

\subsection{Downlink Data Transmission}
Let $s_{k}, \forall k \in \mathcal{K},$ be the symbol for the $k$-th user, where $\mathbb{E}\{|s_k|^2\}=1$ and $\mathbb{E}\{s_is_j^*\}=0$, $\forall i\neq j$, with active beamforming, the transmitted signal from the $b$-th AP is
\begin{equation}
\begin{aligned}\label{eq:transmitted signal}
\mathbf{x}_b &= \sum\limits_{k=1}^{K} \mathbf{f}_{b,k}s_{k},
\end{aligned}
\end{equation}
where $\mathbf{f}_{b,k} \in \mathbb{C}^{N_{\mathrm{t}}}, \forall k \in \mathcal{K},\forall b \in \mathcal{B}$ are active beamforming vectors designed to satisfy the following transmit power constraint at each AP
\begin{equation}\label{eq:transmit power}
\begin{aligned}
\sum\limits_{k=1}^{K}\|\mathbf{f}_{b,k}\|^2 \leq p_{\text{max}},
\end{aligned}
\end{equation}
where $p_{\text{max}}$ is the maximum transmit power and all APs have the same maximum transmit power constraint.

Hence, the received signal at the $k$-th user is expressed as
\begin{equation}
\begin{aligned}\label{eq:received signal}
y_k &= \sum\limits_{b=1}^{B}  \tilde{\mathbf{h}}_{b,k}^H \mathbf{x}_b + n_k =  \sum\limits_{b=1}^{B}  \sum\limits_{j=1}^{K} \tilde{\mathbf{h}}_{b,k}^H\mathbf{f}_{b,j}s_{j}+ n_k,
\end{aligned}
\end{equation}
where $n_k \sim \mathcal{CN}(0, \sigma^2_k)$ indicates the complex Gaussian noise with variance $\sigma_k^2$ at user-$k$.
Then, the signal-to-interference-plus-noise ratio (SINR) of user-$k$ can be given by
\begin{equation}\label{eq:SINR}
\begin{aligned}
\mathrm{SINR}_k &= \frac { \Big| \sum\limits_{b=1}^{B} \tilde{\mathbf{h}}_{b,k}^H \mathbf{f}_{b,k} \Big|^2} { \sum\limits_{j \neq k}^{K} \Big|\sum\limits_{b=1}^{B} \tilde{\mathbf{h}}_{b,k}^H \mathbf{f}_{b,j}\Big|^2 + \sigma_k^2 }.
\end{aligned}
\end{equation}

\subsection{Sum-rate Maximization Problem Formulation}
The goal is to maximize the weighted sum-rate of all $K$ users by designing active beamformer of APs and passive beamformer of RIS elements subject to transmit power constraints in (\ref{eq:transmit power}) and unit-modulus constraints of RIS elements, i.e., $|\theta_m| = 1$.
Thus, the weighted sum-rate maximization problem $\mathcal{P}_o$ is formulated as
\begin{subequations}\label{eq:P0}
\begin{align}
\mathcal{P}_o: \underset{\{\mathbf{f}_{b,k}\}_{\forall b,k}, \bm{\theta} }{\max}~&R_{\mathrm{sum}} =  \sum\limits_{k=1}^{K} \eta_k \mathrm{log}_2(1+\mathrm{SINR}_k) \\
\mathrm{s.t.}~~~&|\theta_m| = 1, ~\forall m,\\
&\sum\limits_{k=1}^{K}\|\mathbf{f}_{b,k}\|^2 \leq p_{\text{max}},~\forall b \in \mathcal{B},
\end{align}
\end{subequations}
where weight coefficient $\eta_k \in \mathbb{R}^+$ indicates the priority of user-$k$.
Obviously, problem $\mathcal{P}_o$ is difficult to optimize due to the non-convexity of objective function (\ref{eq:P0}a) and the unit-modulus constraints (\ref{eq:P0}b).

\section{Partially Distributed Beamforming Design}
The core idea of distributed beamforming design is to decompose centralized problem/constraints into some distributed problems/constraints across APs.
In problem $\mathcal{P}_o$, constraints (\ref{eq:P0}b) and (\ref{eq:P0}c) are distributed, which is beneficial for distributed beamforming design.
However, active beamformer and passive beamformer are coupled with each other in the non-convex objective function (\ref{eq:P0}a).
On the other hand, if fully distributed beamforming design is adopted, the passive beamforming $\bm{\theta}$ used/optimized by each AP should guarantee consensus (i.e., requiring more backhaul signaling among APs) and the large-dimension of $\bm{\theta}$ will make APs undertake a high computational complexity, which is a pyrrhic victory for distributed beamforming schemes.

Motivated by these facts, we propose a partially distributed beamforming design algorithm for solving problem $\mathcal{P}_o$ to avoid additional computational complexity of APs, in which the CPU takes responsibility for large-dimensional passive beamforming optimization and each AP only needs to locally optimize its small-dimensional active beamformer.
In the following, we first transform the weighted sum-rate maximization problem into an equivalent weighted sum-MSE minimization problem \cite{Q. Shi2011}.
Then, AO-based algorithm, a commonly used method in RIS-aided system \cite{Q. Wu2019}, \cite{Z. Zhang2021}, is employed to iteratively optimize active beamforming and passive beamforming between APs and CPU with corresponding approaches.

\subsection{Equivalent Transformation}
In order to make the objective function (\ref{eq:P0}a) convex and realize the equivalent transformation, i.e., from weighted sum-rate maximization problem to weighted sum-MSE minimization problem, we first introduce the mean-square error (MSE) value of the $k$-th user as
\begin{equation}
\begin{aligned}\label{eq:mse_k}
\mathrm{mse}_{k} &\triangleq \mathbb{E} \big\{(u_k^*y_k - s_k)(u_k^*y_k - s_k)^*\big\}, \\
&= u_k^* \Big( \sum\limits_{b=1}^{B}  \sum\limits_{j=1}^{K} \tilde{\mathbf{h}}_{b,k}^H \mathbf{f}_{b,j} \mathbf{f}_{b,j}^H  \tilde{\mathbf{h}}_{b,k} +\sigma_k^2 \Big)u_k  \\
&\hspace{0.4cm}-2 \mathfrak{Re}\Big\{ u_k^* \big(\sum\limits_{b=1}^{B} \tilde{\mathbf{h}}_{b,k}^H \mathbf{f}_{b,k}\big) \Big\}+1,
\end{aligned}
\end{equation}
where $u_k$ is an auxiliary variable for the $k$-th user.
Then, by introducing the weight coefficient $\omega_k$ for each $\mathrm{mse}_k$, problem $\mathcal{P}_o$ can be equivalently transformed into $\mathcal{P}_1$ as \cite{Q. Shi2011}
\begin{equation} \label{eq:P1}
\begin{aligned}
\mathcal{P}_1:\underset{ \mathbf{u}, \bm{\omega}, \bm{\theta}, \{\mathbf{f}_{b,k}\}_{\forall b,k}}{\max} & R_{\mathrm{o}} = \sum\nolimits_{k=1}^{K} \mathrm{ln} (\omega_{k})- \omega_{k} \mathrm{mse}_k+1   \\
\mathrm{s.t.}~~~~&(\ref{eq:P0}\textrm{b}),(\ref{eq:P0}\textrm{c}),
\end{aligned}
\end{equation}
where the weighted sum-MSE (i.e., $\sum_{k=1}^{K} \omega_k\mathrm{mse}_k$) is convex with respect to either the active or the passive beamforming.
Thus, we propose the partially distributed beamforming design algorithm to optimize active beamformer and passive beamformer by alternatively updating variables $\{\mathbf{u}, \bm{\omega}, \bm{\theta}\}$ in CPU and $\{\mathbf{f}_{b,k}\}_{\forall b,k}$ in local APs.
In other words, the partially distributed algorithm is iteratively executed between CPU and APs via the reliable and high-speed backhaul transmission.
In the following, we provide the optimal solutions of active beamforming and passive beamforming in each iteration.

\subsection{Active Beamforming: Fix ($\mathbf{u}$, $\bm{\omega}$, $\bm{\theta}$) and Solve $\mathbf{f}_b^{\text{opt}}$}
In the case of given ($\mathbf{u}$, $\bm{\omega}$, $\bm{\theta}$), by removing constant term, the equivalent problem $\mathcal{P}_1$ can be recast as the following sub-problem $\mathcal{P}_{\text{active}}$ for active beamforming design
\begin{equation}
\begin{aligned}\label{eq:P solve f}
\mathcal{P}_{\text{active}}:\underset{\{\mathbf{f}_{b,k}\}_{\forall b,k}}{\min}&R_{\mathrm{a}}(\mathbf{f}_{b,k}) = \sum\nolimits_{k=1}^{K} \omega_k \mathrm{mse}_k \\
\mathrm{s.t.}~~~&(\ref{eq:P0}\textrm{c}).
\end{aligned}
\end{equation}

\nid Although the weighted sum-MSE objective function in (\ref{eq:P solve f}) is convex with respect to the active beamforming $\mathbf{f}_{b,k},\forall b,k$, $\mathcal{P}_{\text{active}}$ is still a centralized problem.
In order to realize the distributed active beamforming design, we decompose problem $\mathcal{P}_{\text{active}}$ into $B$ distributed sub-problems and reformulate the objective function in (\ref{eq:P solve f}) as
\begin{equation}\label{eq:function active reform}
\begin{aligned}
R_{\tilde{\mathrm{a}}}(\mathbf{f}_b) = \sum_{b = 1}^{B}  \big( \mathbf{f}_b^H \mathbf{W}_b \mathbf{f}_b - 2 \mathfrak{Re} \{ \mathbf{v}_b^H \mathbf{f}_b \} \big)+ c,
\end{aligned}
\end{equation}
where $\mathbf{f}_b \triangleq [\mathbf{f}_{b,1}^T, \mathbf{f}_{b,2}^T, \ldots, \mathbf{f}_{b,K}^T]^T \in \mathbb{C}^{N_{\mathrm{t}}K}$ denotes the active beamforming vector of the $b$-th AP and the following matrices/vectors are defined for brevity:
\begin{subequations}
\begin{align}
&\mathbf{W}_b\triangleq  \mathbf{I}_{K} \otimes  \Big\{ \sum\nolimits_{k=1}^{K} \omega_k |u_k|^2 \tilde{\mathbf{h}}_{b,k}  \tilde{\mathbf{h}}_{b,k}^H  \Big\},\\
&c\triangleq \sum\nolimits_{k=1}^{K} \omega_k(1 + |u_k|^2 \sigma_k^2), \\
&\mathbf{v}_{b}\triangleq [\mathbf{v}_{b,1}^T, \mathbf{v}_{b,2}^T, \ldots, \mathbf{v}_{b,K}^T], \\
&\mathbf{v}_{b,k}\triangleq  \omega_k u_k \tilde{\mathbf{h}}_{b,k}.
\end{align}
\end{subequations}

\nid Obviously, function (\ref{eq:function active reform}) can be equivalently divided into $B$ sub-functions across APs after dropping constant term $\textit{c}$, i.e., $\min(\sum_{b=1}^{B} f_b) =  \sum_{b=1}^{B} (\min f_b)$.
Hence, the sub-problem $\mathcal{P}_{\text{active}}^{(b)}$ of the $b$-th AP is given by
\begin{equation}
\begin{aligned}\label{eq:P solve f2}
\mathcal{P}_{\text{active}}^{(b)}:\underset{\mathbf{f}_b}{\min}~~~&  \mathbf{f}_b^H \mathbf{W}_b \mathbf{f}_b - 2 \mathfrak{Re} \{ \mathbf{v}_b^H \mathbf{f}_b \}\\
\mathrm{s.t.}~~~&\mathbf{f}_b^H \mathbf{f}_b \leq p_{\text{max}}.
\end{aligned}
\end{equation}
Since the matrices $\mathbf{W}_b$ is positive semidefinite, problem $\mathcal{P}_{\text{active}}^{(b)}$ can be solved by many methods, such as primal-dual sub-gradient \cite{Z. Zhang2021}, ADMM \cite{S. Huang2021}, etc.
One of the outstanding advantages of cell-free networks is the larger number of APs, but the smaller number of antennas per AP.
Therefore, we derive the closed-form of active beamformer for all APs and the optimal solution of $\mathbf{f}^{\text{opt}}_{b}$ for problem $\mathcal{P}_{\text{active}}^{(b)}$ is
\begin{equation}
\begin{aligned}\label{eq:solution of f}
\mathbf{f}^{\text{opt}}_{b} = (\mathbf{W}_b + \lambda_b \mathbf{I}_{N_{\mathrm{t}}K})^{-1}  \mathbf{v}_b,
\end{aligned}
\end{equation}
where $\{\lambda_b\}_{b \in \mathcal{B}}$ is the introduced multiplier for transmit power constraints, which can be obtained by bisection search method.

\subsection{Passive Beamforming: Fix $\mathbf{f}_b$ and Solve ($\mathbf{u}^{\text{opt}}$, $\bm{\omega}^{\text{opt}}$, $\bm{\theta}^{\text{opt}}$)}
With the obtained active beamformers $\{\mathbf{f}_{b}\}_{b \in \mathcal{B}}$ from all APs, the sub-problem for passive beamforming design is
\begin{equation}\label{eq:P solve theta1}
\begin{aligned}
\mathcal{P}_{\text{passive}}:\underset{\mathbf{u}, \bm{\omega}, \bm{\theta}}{\max}~~&\sum_{k=1}^{K} \mathrm{ln} (\omega_{k})- \omega_{k} \mathrm{mse}_k+1   \\
\mathrm{s.t.}~~& (\ref{eq:P0}\textrm{b}).
\end{aligned}
\end{equation}

\nid In the following, we derive the closed-form solutions of variables $\{\mathbf{u}, \bm{\omega}\}$, relax the non-convex constraints \cite{C. Pan2021}, and then optimize passive beamformer $\bm{\theta}$.

\textit{i}): Update auxiliary variable $\mathbf{u}^{\text{opt}}$. When variables $\bm{\omega}$, $\bm{\theta}$, and $\{\mathbf{f}_{b}\}_{b \in \mathcal{B}}$ are fixed, the objective function in (\ref{eq:P solve theta1}) is concave with respect to each $u_k$ term.
Therefore, the optimal solution of $\mathbf{u}^{\text{opt}}$ for problem $\mathcal{P}_{\text{passive}}$ can be obtained by setting $\{\partial \omega_k\mathrm{mse}_k /\partial u_k \}_{\forall k \in \mathcal{K}}$ to zero, which is given by
\begin{equation}\label{eq:u}
\begin{aligned}
u_{k}^{\text{opt}}= \frac{ \sum\limits_{b=1}^{B} \tilde{\mathbf{h}}_{b,k}^H \mathbf{f}_{b,k}}{\sum\limits_{b=1}^{B}  \sum\limits_{j=1}^{K} \tilde{\mathbf{h}}_{b,k}^H \mathbf{f}_{b,j} \mathbf{f}_{b,j}^H  \tilde{\mathbf{h}}_{b,k} +\sigma_k^2 }.
\end{aligned}
\end{equation}

\textit{ii}): Update weight coefficient $\bm{\omega}^{\text{opt}}$. The sub-problem of $\bm{\omega}$ is that ${\bm{\omega}^{\text{opt}}} = {\arg} ~{\max} ~\sum_{k=1}^{K} \mathrm{ln} (\omega_{k})- \omega_{k} \mathrm{mse}_k+1$, whose optimal solution is $\omega_{k}=\mathrm{mse}_k^{-1}$,~$\forall k \in \mathcal{K}$.
Moreover, the $\mathrm{mse}_k$ for each user-$k$ of problem $\mathcal{P}_{\text{passive}}$ can be easily calculated by substituting (\ref{eq:u}) into equation (\ref{eq:mse_k}), which is expressed as
\begin{equation}\label{eq:mseopt}
\begin{aligned}
\mathrm{mse}_k^{\text{opt}}= 1 - \frac{ \sum\limits_{b=1}^{B} |\tilde{\mathbf{h}}_{b,k}^H \mathbf{f}_{b,k}|^2 }{\sum\limits_{b=1}^{B}  \sum\limits_{j=1}^{K}| \tilde{\mathbf{h}}_{b,k}^H \mathbf{f}_{b,j}|^2 +\sigma_k^2 }.
\end{aligned}
\end{equation}

\textit{iii}): Update passive beamformer $\bm{\theta}$. With the obtained solutions of $\{\mathbf{u}^{\text{opt}}, \bm{\omega}^{\text{opt}}\}$ and removing constant term, the sub-problem of passive beamforming design can be recast as
\begin{equation}\label{eq:P solve theta2}
\begin{aligned}
\mathcal{P}_{\text{passive}}^{'}:\underset{\bm{\theta}}{\min}~~~ &\bm{\theta}^H \mathbf{Q} \bm{\theta} - 2 \mathfrak{Re} \{ \mathbf{p}^H \bm{\theta} \}\\
\mathrm{s.t.}~~~&(\ref{eq:P0}\textrm{b}),
\end{aligned}
\end{equation}
where the following matrices/vectors are defined for brevity:
\begin{subequations}
\begin{align}
\mathbf{Q} &\triangleq \sum_{k =1}^{K}\sum_{b =1}^{B} \sum_{j =1}^{K}\omega_k |u_k\mathbf{q}_{b,k}^H \mathbf{f}_{b,j}|^2,\\
\mathbf{p} &\triangleq  \sum_{k =1}^{K} \sum_{b =1}^{B} \big(\omega_k u_k^* \mathbf{q}_{b,k}^H \mathbf{f}_{b,k} - \sum_{j =1}^{K}\omega_k |u_k|^2 \mathbf{q}_{b,k}^H \mathbf{f}_{b,j} \mathbf{f}_{b,j}^H \mathbf{h}_{b,k}\big).
\end{align}
\end{subequations}
Note that $\mathbf{q}_{b,k}^H \triangleq \mathrm{diag}(\mathbf{v}_{k}^H) \mathbf{G}_{b}, \forall b,k,$ is the required CSIs from APs.
We relax the non-convex constraint (\ref{eq:P0}\textrm{b}) to the convex constraint $\tilde{\mathcal{S}_1} = \{\bm{\theta} | |\theta_{m}| \leq 1, \theta_{m} \in \mathbb{C}, \forall m \}$ with which problem $\mathcal{P}_{\text{passive}}^{'}$ can be easily solved by the standard convex tools, e.g., CVX \cite{C. Pan2021}.
Benefit from the re-arrangement of objective function in (\ref{eq:P solve theta2}), passive beamforming design can also be optimized by other algorithms (i.e., majorization-minimization (MM) algorithm and manifold approach) with the help of strong centralized processing ability of CPU.

\subsection{Algorithm Implementation and Analysis}
The partially distributed beamforming design algorithm is implemented as follows.
First, AP-$b$ estimates CSIs (i.e., $\{\mathbf{h}_{b,k}\}_{\forall k}$, $\mathbf{G}_{b}$, and $\{\mathbf{v}_{k}\}_{\forall k}$) to obtain the equivalent baseband channel $\{\tilde{\mathbf{h}}_{b,k}\}_{\forall k}$ and then share required information (i.e., $\{\mathbf{h}_{b,k}\}_{\forall k}$ and $\{\mathbf{q}_{b,k}\}_{\forall k}$) with CPU.
In each iteration, AP-$b$ begin with initializes $\{\mathbf{u}^{(i)}, \bm{\omega}^{(i)}, \bm{\theta}^{(i)}\}$ (or receives them from CPU) and locally optimizes active beamformer $\mathbf{f}_{b}^{(i+1)}$.
Next, all APs feed back $\mathbf{f}_{b}^{(i+1)}$ to CPU.
With the obtained active beamformer, CPU optimizes the introduced variable $\mathbf{u}^{(i+1)}$, weight coefficient $\bm{\omega}^{(i+1)}$, passive beamformer $\bm{\theta}^{(i+1)}$.
Then, the CPU returns $\{\mathbf{u}^{(i+1)}, \bm{\omega}^{(i+1)}, \bm{\theta}^{(i+1)}\}$ to all APs for next iteration.
In conclusion, the partially distributed beamforming design algorithm is summarized in $\textbf{Algorithm 1}$.
Next, we provide a concise signaling overhead analysis.
The CSIs exchange requires $2BN_{\mathrm{t}}K$ backhaul signaling.
According to steps 4 and 7 in Algorithm 1, the required signaling for updating in each AP and CPU are $(M+2K)$ and $BN_{\mathrm{t}}K$ symbols, respectively.
The total required signaling of the proposed method is $2BN_{\mathrm{t}}K + I_i(M+2K+BN_{\mathrm{t}}K)$ symbols, where in $I_i$ denotes the number of iterations.

\begin{algorithm}[t]
\begin{small}
  \caption{\small{Partially Distributed Beamforming Design Algorithm}}
  \label{alg:Algorithm 2}
  \begin{algorithmic}[1]
    \REQUIRE $B, N_{\textrm{t}}, K, M$.
    \ENSURE Active beamformer $\{\mathbf{f}_b\}_{b \in \mathcal{B}}$ and passive beamformer $\bm{\theta}$.
    \STATE{The APs estimate CSIs (i.e., $\{\mathbf{h}_{b,k}\}_{\forall b,k}$, $\{\mathbf{v}_{k}\}_{\forall k}$, and $\{\mathbf{G}_{b}\}_{\forall b}$) and then feed back $\{\mathbf{h}_{b,k}\}_{\forall b,k}$ and $\{\mathbf{q}_{b,k}\}_{\forall b,k}$ to CPU.}
    \WHILE{no convergence of $R_{\mathrm{sum}}$}
    \STATE{AP side ($\forall b$):}
    \STATE{~~~Initialize/receive $\mathbf{u}^{(i-1)}, \bm{\omega}^{(i-1)}, \bm{\theta}^{(i-1)}$;}
    \STATE{~~~Update $\mathbf{f}_b^{(i)}$ by solving $\mathcal{P}_{\text{active}}^{(b)}$ with (\ref{eq:solution of f});}
    \STATE{CPU side:}
    \STATE{~~~Receive $\mathbf{f}_b^{(i)}$ from all APs;}
    \STATE{~~~Update $\mathbf{u}^{(i)}$ and $\bm{\omega}^{(i)}$ with (\ref{eq:u}) and (\ref{eq:mseopt});}
    \STATE{~~~Update $\bm{\theta}^{(i)}$ by solving $\mathcal{P}_{\text{passive}}^{'}$ in (\ref{eq:P solve theta2});}
    \STATE{\textit{i} := \textit{i} + 1;}
    \ENDWHILE
  \end{algorithmic}
\end{small}
\end{algorithm}

Then, we analyze the convergence and computational complexity.
The optimal solutions of $\mathbf{f}_{b}$, $\mathbf{u}$, and $\bm{\omega}$ are provided in the closed-form expression. The updating of $\bm{\theta}$ by the standard convex tools maximizes the objective function. Therefore, the equivalent objective function $R_{\mathrm{o}}$ is monotonically nondecreasing after each iteration and the proposed algorithm converges to at least a local optimum \cite{Q. Shi2011}.
Next, in each iteration, the computational complexity is mainly caused by the matrix inversion in (\ref{eq:solution of f}) and the optimization of $\bm{\theta}$, which lead to the complexity of $\mathcal{O}\{(N_{\mathrm{t}}K)^3\}$ and $\mathcal{O}(M^{3.5})$, respectively.
The total computational complexity is $\mathcal{O}\{I_i(M^{3.5} +B(N_{\mathrm{t}}K)^3) \}$.

To sum up, the proposed partially algorithm realizes signaling overhead reduction compared with the fully distributed cooperative beamforming method, which has $B^2(N_{\mathrm{t}}K + I_i(N_{\mathrm{t}}K +M +2K))$ signaling overhead.
Meanwhile, the proposed algorithm reduces the computational complexity of CPU by dispersing active beamformer $\mathbf{f}$ into $\{\mathbf{f}_b\}_{\forall b}$ and ensures the consistency and accuracy of passive beamforming design. Besides, reliable backhaul transmission between APs and CPU makes the proposed partially distributed beamforming design algorithm more practical compared with other uplink-downlink iteration algorithms, e.g., \cite{I. Atzeni2021}.

\section{Simulation Results}
\begin{figure*}[t]
\centering
  \begin{minipage}{0.329\linewidth}
  \centering
  \includegraphics[width = \linewidth]{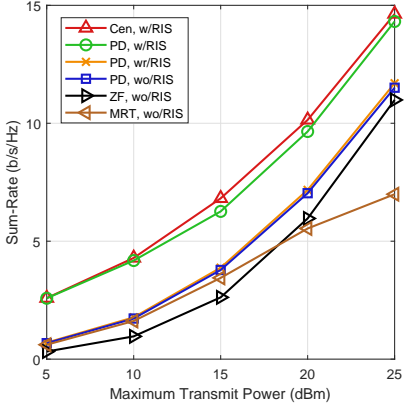}
  \vspace{-0.4 cm}
  \caption{Sum-rate versus transmit power.}
  \label{fig:Power_Vs_Sumrate}
  \end{minipage}
  \begin{minipage}{0.329\linewidth}
  \centering
  \includegraphics[width= \linewidth]{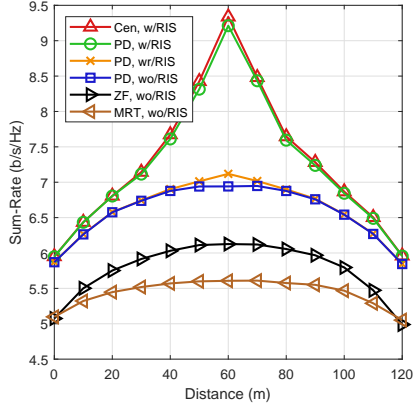}
  \vspace{-0.4 cm}
  \caption{Sum-rate versus user's location.}
  \label{fig:Distance_Vs_Sumrate}
  \end{minipage}
  \begin{minipage}{0.329\linewidth}
  \centering
  \includegraphics[width= \linewidth]{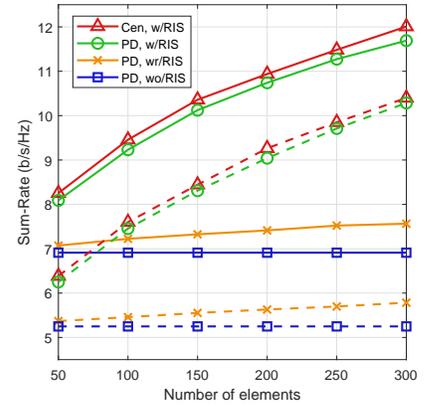}
  \vspace{-0.4 cm}
  \caption{Sum-rate versus RIS elements.}
  \label{fig:elements_Vs_Sumrate}
  \end{minipage}
  \vspace{0.2 cm}
\end{figure*}

In this section, simulation results are provided to evaluate the performance of proposed partially distributed beamforming design algorithm for RIS-aided cell-free networks.
$B = 5$ APs are equipped with $N_{\mathrm{t}} = 8$ antennas and located at (0, -50 m), (30 m, -50 m), (60 m, -50 m), (90 m, -50 m), and (120 m, -50 m), respectively.
$K = 4$ single-antenna users are randomly distributed in a circle centered at (60 m, 0) with radius 5 m.
An RIS is located at (60 m, 10 m) and consists of $M = 100$ reflecting elements.
Besides, we adopt the commonly used large-scale fading model and the distance-dependent path loss model $L(d) = C_0 (d/d_0)^{-\kappa}$, where $C_0$ = -32 \text{dB} is the path loss as the reference distance $d_0$ = 1 \text{m} and $\kappa$ is the pathloss exponent \cite{Q. Wu2019}, \cite{Z. Zhang2021}.
The AP-RIS link is line-of-sight (LOS) modeled by Rayleigh fading channel.
The AP-user link and RIS-user link are non-line-of-sight (NLOS) modeled by Rayleigh fading channel \cite{S. Huang2021}.
Then, the pathloss exponents of AP-user link, AP-RIS link, and RIS-user link are set as 3.6, 2.2, and 2.6, respectively.
Other system parameters are set as follows: $\sigma_k^2  = -70$ dBm, $\forall k$, $\epsilon = 10^{-3}$, and $\eta_k = 1$,$\forall k$.

In Fig. \ref{fig:Power_Vs_Sumrate}, we evaluate the sum-rate performance of proposed partially distributed beamforming design algorithm with RIS ($\textbf{PD, w/RIS}$) with respect to the maximum transmit power $p_{\mathrm{max}}$.
We also include the following algorithms: \textit{i}) Centralized beamforming design with RIS ($\textbf{Cen, w/RIS}$), which serves performance upper bound;
\textit{ii}) Partially distributed beamforming design with random RIS ($\textbf{PD, wr/RIS}$), the passive beamformer $\bm{\theta}$ is randomly selected and not optimized;
\textit{iii}) Partially distributed beamforming design without RIS ($\textbf{PD, wo/RIS}$);
\textit{iv}) Fully distributed local ZF beamforming design without RIS ($\textbf{ZF, wo/RIS}$);
\textit{v}) Fully distributed MRT beamforming design without RIS ($\textbf{MRT, wo/RIS}$).
It can be observed that the proposed algorithm achieves the performance close to the centralized beamforming scheme and always outperforms the conventional local beamforming methods.
In addition, the considerable performance gap between the proposed algorithm with RIS and random/without RIS further indicates the promising potential of deploying RIS in cell-free networks.

Fig. \ref{fig:Distance_Vs_Sumrate} shows the sum-rate as a function of user's location $D_{\textrm{u}}$, where $p_{\mathrm{max}}=20$ dBm.
The similar conclusion can be noticed that the proposed partially algorithm can obtain near-optimal performance.
Moreover, there is a performance peak at $D_{\textrm{u}}$ = 60 m, which indicates that sum-rate increases when users approach the RIS.
However, the peak does not appear for the conventional local beamforming methods without RIS and it further illustrates that the capacity of cell-free networks can be substantially increased by deploying RIS.

Fig. \ref{fig:elements_Vs_Sumrate} illustrates the sum-rate against the number of RIS elements $M$.
The solid lines and dotted lines illustrate the performances for the systems with 4 users and 2 users, respectively.
We can observe that sum-rate performance of RIS-aided cell-free networks increases with more RIS elements.
More importantly, with the increasing number of RIS elements, the gap between proposed partially distributed beamforming algorithm with RIS and random RIS is gradually larger, which confirms the necessity of passive beamforming design for RIS.

\section{Conclusions}
In this paper, we studied the weighted sum-rate maximization problem and proposed a partially distributed beamforming design algorithm for RIS-aided cell-free networks.
The sum-rate maximization problem is first transformed into an equivalent weighted sum-MSE minimization problem and then the AO-based approach is adopted to iteratively optimize active beamformer and passive beamformer between APs and CPU.
The closed-form of active beamformers are obtained locally by APs and the passive beamformer is centralized designed by CPU.
Numerical results illustrated the proposed algorithm can achieve a remarkable performance improvement compared with other local beamforming methods as well as demonstrated the significant potential of deploying RIS in cell-free networks.


\begin{thebibliography}{99}

\bibitem{E. Bjornson2017}  E. Bj\"{o}rnson, J. Hoydis, and L. Sanguinetti, ``Massive MIMO networks: Spectral, energy, and hardware efficiency,'' \textit{Found. Trends Signal Process.}, vol. 11, no. 3-4, pp. 154-655, Nov. 2017.

\bibitem{H. Q. Ngo2017}   H. Q. Ngo, A. Ashikhmin, H. Yang, E. G. Larsson, and T. L. Marzetta, ``Cell-free massive MIMO versus small cells,'' \textit{IEEE Trans. Wireless Commun.}, vol. 16, no. 3, pp. 1834-1850, Mar. 2017.

\bibitem{O. T. Demir2021}   \"{O}. T. Demir, E. Bj\"{o}rnson, and L. Sanguinetti, ``Foundations of user-centric cell-free massive MIMO,'' \textit{Found. Trends Signal Process.}, vol. 14, no. 3-4, pp. 162-472, Jan. 2021.

\bibitem{Y. Zhang2019}    Y. Zhang, H. Cao, M. Zhou, and L. Yang, ``Cell-free massive MIMO: Zero forcing and conjugate beamforming receivers,'' \textit{J. Commun. Networks}, vol. 21, no. 6, pp. 529-538, Dec. 2019.

\bibitem{E. Bjornson2020}  E. Bj\"{o}rnson and L. Sanguinetti, ``Making cell-free massive MIMO competitive with MMSE processing and centralized implementation,'' \textit{IEEE Trans. Wireless Commun.}, vol. 19, no. 1, pp. 77-90, Jan. 2020.

\bibitem{I. Atzeni2021} I. Atzeni, B. Gouda, and A. T\"{o}lli, ``Distributed precoding design via over-the-air signaling for cell-free massive MIMO,'' \textit{IEEE Trans. Wireless Commun.}, vol. 20, no. 2, pp. 1201-1216, Feb. 2021.

\bibitem{Q. Wu2019}    Q. Wu and R. Zhang, ``Intelligent reflecting surface enhanced wireless network via joint active and passive beamforming,'' \textit{IEEE Trans. Wireless Commun.}, vol. 18, no. 11, pp. 5394-5409, Nov. 2019.


\bibitem{X. Zhou2022} X. Zhou, S. Yan, Q. Wu, F. Shu, and D. W. K. Ng, ``Intelligent reflecting surface (IRS)-aided covert wireless communications with delay constraint,"  \textit{IEEE Trans. Wireless Commun.}, vol. 21, no. 1, pp. 532-547, Jan. 2022.

\bibitem{C. Pan2021}   C. Pan, \textit{et al.}, ``An overview of signal processing techniques for RIS/IRS-aided wireless systems,'' \textit{IEEE J. Selected Topics Signal Process.}, early access.

\bibitem{Z. Zhang2021}     Z. Zhang and L. Dai, ``A joint precoding framework for wideband reconfigurable intelligent surface-aided cell-free network,'' \textit{IEEE Trans. Signal Process.}, vol. 69, pp. 4085-4101, Sep. 2021.

\bibitem{S. Huang2021}    S. Huang, Y. Ye, M. Xiao, H. V. Poor, and M. Skoglund, ``Decentralized beamforming design for intelligent reflecting surface-enhanced cell-free networks,'' \textit{IEEE Wireless Commun. Lett.}, vol. 10, no. 3, pp. 673-677, Mar. 2021.

\bibitem{Q. Shi2011}   Q. Shi, M. Razaviyayn, Z. Luo, and C. He, ``An iteratively weighted MMSE approach to distributed sum-utility maximization for a MIMO interfering broadcast channel,'' \textit{IEEE Trans. Signal Process.}, vol. 59, no. 9, pp. 4331-4340, Sep. 2011.


\end{thebibliography}
\end{document}